# Social media attention increases article visits: An investigation on article-level referral data of *PeerJ*

Xianwen Wang*, Yunxue Cui, Qingchun Li, Xinhui Guo

WISE Lab, Faculty of Humanities and Social Sciences, Dalian University of Technology, Dalian 116085, China.

* Corresponding author.
Email address: xianwenwang@dlut.edu.cn; xwang.dlut@gmail.com

**Abstract**: In order to better understand the effect of social media in the dissemination of scholarly articles, employing the daily updated referral data of 110 *PeerJ* articles collected over a period of 345 days, we analyze the relationship between social media attention and article visitors directed by social media. Our results show that the social media presence of *PeerJ* articles is high. About 68.18% of the papers receive at least one tweet from Twitter accounts other than @PeerJ, the official account of the journal. Social media attention increases the dissemination of scholarly articles. Altmetrics could not only act as the complement of traditional citation measures, but also play an important role in increasing the article downloads and promoting the impacts of scholarly articles. There also exist significant correlation among the online attention from different social media platforms. Articles with more Facebook shares tend to get more tweets. The temporal trends show that social attention comes immediately following publication but does not last long, so do the social media directed article views.
**Keywords**: Altmetrics, social media, Twitter, PeerJ, referral

## 1. Introduction
### 1.1 Social media attention about scholarly articles
Social media, such as Facebook and Twitter, has become a critical tool in scholarly communications. Dissemination of research in traditional way depends on the user searching for or 'pulling' relevant knowledge from the literature base. Social media, instead, 'pushes' knowledge to the user straightly (Allen et al. 2013). Not only general public, scientists are also active users of social media (Rowlands et al. 2011; Van Noorden, 2014; Veletsianos 2016). According to the estimation of altmetric.com, around 15,000 unique research outputs are shared or mentioned online each day (Altmetric 2016). About 21.5% of papers receive at least one tweet overall, however, Twitter density is very different in different fields, higher in Social Sciences, Biomedical and Health Sciences, as well as Life and Earth Sciences, but very low in Mathematics and Computer Science and Natural Sciences and Engineering (Haustein, Costas, & Larivière, 2015). Open access is also an important factor in disseminating articles on social media. Open access articles receive more social media attention and higher article downloads than non-open access papers (Wang et al. 2015).

*1.2 Relationship among social media attention, downloads and citation*

Altmetrics can supply hints of online concerns from publics. Moreover, altmetrics is in some measure produced by scholars as part of their academic communication (Lăzăroiu, 2017). The correlations between altmetrics and citation are complicated. Firstly, there are correlations between Mendeley readership and times cited (Priem, Piwowar, & Hemminger, 2012; Zahedi, Costas, & Wouters, 2017). Nevertheless, the results of correlations between tweets and citation are controversy. For example, an early study confirms that tweets can predict highly cited articles within the first 3 days after article publication (Eysenbach, 2011); however, some other studies draw different conclusions. Thelwall, et al (2013) found that tweets are associated with citation counts, but there is no correlation between altmetrics and citations. Haustein, et al. (2014) and Costas, et al. (2015) found very weak correlation between the number of tweets and the number of citations of papers. Thirdly, there are positive correlations between downloads and citations, most downloaded articles are those that are more likely to receive citations (O'Leary, 2008; Lippi & Favaloro, 2012). Article download is one of the first alternative metrics to be introduced in digital library (Bollen et al, 2009; Kurtz & Bollen. 2010), while link analysis at article-level is even earlier to be used as altmetrics indicators for research evaluation (Kousha & Thelwall. 2007a; Kousha & Thelwall. 2007b) As early as 2004, the BMJ provided the article views to public. Nowadays, article usage data are available on the article page from a lot of publishers' and individual journals' websites, including Springer Nature, Frontiers, IEEE Xplore Digital Library, ACM Digital Library, Taylor & Francis, Oxford University Press, *Science, PNAS* and *PeerJ*, etc. (Wang et al. 2014; Wang, Fang & Sun, 2016). The blooming of usage data inspires many studies from various perspectives, i.e., exploring researchers' working habits according to the time of article downloads (Wang et al. 2012),the temporal trends of article downloads after publication (Wang et al. 2014; Khan & Younas. 2017; Duan & Xiong. 2017). Compared to downloads, citations usually delay by about 2 years, so download statistics provide a useful indicator of eventual citations in advance (Watson, 2009). More downloads during a limited time period is an indicator of more citations to the article in a long-term interval (Jahandideh, Abdolmaleki, & Asadabadi, 2007). Yan and Gerstein (2011) found that there are intrinsic differences among different types of article usage (HTML views and PDF downloads versus XML). PDF downloads increase the probability that people would later read it (Allen, Stanton, Pietro, & Moseley, 2013). The fourth aspect is concerning the relationship between social media attention and article downloads. It is considered that people hardly read the articles they tweet about, for example, Haile (2014) stated that they "found effectively no correlation between social shares and people actually reading". Employing a small dataset (16 articles), Allen et al. (2013) reported that social media release of a research article in the clinical pain sciences increases the article visitors. In our previous study, applying the referral data from *PeerJ*, we found that referrals from social media account for a significant number of visits to articles, especially during the days shortly after publication. However, this fast initial accumulation soon gives way to a rapid decay (Wang et al., 2016). Winter (2015) found a clear association exists between the number of tweets and the number of views for PLOS ONE articles.

It is necessary to point out that article-level metrics is different from author-level metrics (ALMetrics) within altmetrics, where the latter measure the impact of individual authors through varied metric indicators, including bibliometrics, usage, participation, rating, social connectivity, and composite indicators (Torres-Salinas & Milanes-Guisado, 2014; Orduña-Malea et al, 2016).

### 1.3 Adoption of altmetrics

There are three major services calculating altmetrics, including Altmetric.com, Plum Analytics and Impactstory. Plum Analytics has covered the most number of papers. According to the statistics, it covers 52.6 million research outputs, of which 56.6% (29.7 million) are articles (http://plumanalytics.com/learn/about-metrics/coverage/). Altmetric.com covers over five million research outputs and ImpactStory tracks around 1 million publications.

Table 1 Coverage of major altmetrics services

| Altmetrics service | Coverage |
| --- | --- |
| Plum Analytics | ~ 52.6 million artifacts, 56.6% (29.7 million) are articles[1] |
| Altmetric.com | > 5 million research outputs[2] |
| ImpactStory | ~ 1 million publications[3] |

The importance of social media in disseminating scholarly articles has been realized by publishers. Nowadays, almost all publishers have integrated the social share tools into article page, which makes article readers share articles on social media platforms easily. As two pioneers, Journal of Medical Internet Research (in 2008) and PLoS (in 2009) started to systematically collect tweets about their articles. Now, many publishers have started providing altmetrics statistics to readers. In 2017, PlumX from Plum Analytics is integrated into Scopus. According to the information released by altmetric.com, over 70 publishers now display Altmetric data across their article pages, including Springer Nature, Wiley, Frontiers, and PeerJ, etc.

### 1.4 Research gap and research questions

Previous studies confirmed the correlation between downloads and citations, and the correlation between Mendeley readership and citations, and although some previous studies confirmed the overall association between tweets and article views using the data of tweets and total views of articles, there lacks direct evidence. If we may know the number of tweets about an article and get the tweet directed article views, not only the total article views, we could confirm the causal relationship from the social media attention to the directed article views.

---

[1] 56.6% of 52.6 million artifacts are articles, http://plumanalytics.com/wp-content/uploads/2016/11/Plum-Analytics-Coverage-Infographic.pdf, retrieved 3 August 2017.

[2] https://figshare.com/articles/Altmetric_the_story_so_far/2812843, retrieved 3 August 2017.

[3] https://twitter.com/Impactstory/status/731258457618157568, retrieved 3 August 2017.

Table 1 Research gap and our research questions

| Relationship | Results |
| --- | --- |
| Correlation between downloads and citations | Positive, significant |
| Correlation between social attention and citation | Positive, significant (Mendeley readership and citation); Not significant (Tweets and citation) |
| Correlation between social attention and downloads | To be confirmed in this research |

In this study, with the availability of referrals data at article-level, which will be introduced in the following method part, we are able to examine this kind of causal relationship. Our research questions are, firstly, what is the relationship between social media attention and article views? Does more social media attention suggest more article visitors? Secondly, what is the relationship between different kinds of social media attention? Does the number of tweets of articles associate with activity on other social media?

Answering these questions will validate the effects of social media in promoting the impacts of scholarly articles and shed light on the mechanism of altmetrics in scholarly communication.

## 2. Method

*PeerJ*, an open access, peer reviewed scholarly journal, provides data on the referral source of article visitors to all *PeerJ* article pages, as shown in Figure 1. This is unique because such data is not available on other publishers or journal websites. Although Frontiers also provides partial referral data of each Frontiers article, it only includes the top five referring sites; however, there are usually hundreds of referrals for one paper, so only the data of five referrals is far not sufficient for study. The metrics of *PeerJ* provide all referrals of each paper, no matter how many referrals it has, and update daily since the following day of an article's publication, meaning that we are able to track the digital footprints of scholarly articles.

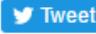

Figure 1 Referrals of a PeerJ paper

The metadata and article visits data are collected from peerj.com directedly, while the data of Tweets and Facebook shares for each article are collected from Plum analytics. We use the same dataset as our previous study (Wang et al, 2016).Because we are studying the temporal trend of article visits since the first day of publication, so a long study period is not appropriate. We choose articles published during the period from January 21, 2016 to February 18, 2016 are selected as the research objects, there are a total of 110 samples included, which accounts for about 6.5% of all *PeerJ* papers up to then. Although the dataset includes only a small section of the total papers, it covers all the main subjects of the journal, which made it an enough fraction of the journal. The referral data are collected and updated daily. Compared with the 90 days of time window used in our previous study (Wang et al, 2016), the time window of this study is extended to 345 days, which covers the date from 22 January, 2016 to December 31, 2016.

The altmetrics data (social media attention) are retrieved from Plum Analytics, which has been integrated into Scopus now, including Tweets and Facebook shares for each article. Here we collect the Plum Analytics data from Scopus manually.

Finally, the metadata, referral data, and altmetrics data are processed and parsed into our designed SQL database for analysis.

In this study, we use statistical methods including correlation analysis and one-way ANOVA. Correlation analysis is used to examine the relationship between social media attention and the number of social media directed visitors, and the relationship between attentions from different social media platforms, etc. One-way ANOVA is used to test whether there are significant differences in the number of tweets and Twitter directed visitors between the two periods.

## 3. Results
### 3.1 Descriptive analysis

All the papers have received at least two tweets. The median total tweets are 6, and median Twitter directed visitors is 11.5, as Table 2 shows. However, since @thePeerJ, the official Twitter account tweets each article twice, on the exact day and the following day of the article publication. If we exclude the tweets from @thePeerJ, the results would be a little different, that is about 68.18% of the papers receive at least one tweet. The median of total tweets is also six, and the median of Twitter directed visitors is 10.5.

Table 2 Statistical results

|  | Tweets | Retweets | Total tweets | Twitter directed visitors |
|---|---|---|---|---|
| Max | 34 | 80 | 100 | 918 |
| Min | 2 | 0 | 2 | 0 |
| Median | 3.5 | 2.5 | 6 | 11.5 |

The most shared paper on Facebook is the article *The furculae of the dromaeosaurid dinosaur Dakotaraptor steini are trionychid turtle entoplastra* (https://doi.org/10.7717/peerj.1691), which is a study about archaeology. It got 196 Facebook shares and also got 63 tweets (ranked 6th among all papers). The most tweeted paper is the article *Evaluation of the global impacts of mitigation on persistent, bioaccumulative and toxic pollutants in marine fish* (https://peerj.com/articles/1573/), which is a study about marine environment protection. It got 100 tweets and also got 49 Facebook shares (ranked 20th among all papers). The most visited paper is the article *The effect of habitual and experimental antiperspirant and deodorant product use on the armpit microbiome* (https://peerj.com/articles/1605/), which is a study about personal health. It got 105 Facebook shares (ranked 8th among all papers) and 97 tweets (ranked 2nd among all papers). In general, most of those top shared and tweeted articles are studies concerning issues include health, animals, and environment, etc.

### 3.2 Correlation analysis
*Correlation between total visitors and visitors directed from social referrals*

Since the data distribution is positively skewed, we use Log transformation. After log transformation, the data (including the data in Figure 3 and 4) obey normal distribution, which is tested by Shapiro-Wilks test. Figure 2 shows the relationship between visitors directed from social referrals and total article visitors with log transformation as of December 31, 2016. Because Spearman correlation test does not assume any assumptions about the distribution of the data and is the appropriate correlation analysis when the variables are measured on a scale that is at least ordinal, so we adopt Spearman correlation analysis in this research. The result indicates that exists a positive and strong association between the two variables. Social media mentions are positively and strongly correlated with the resulted article visits, while the correlation coefficient r =

0.785 (p<0.001). In other words, the more social media mentions an article receives, the more visitors it attracts from social media referrals.

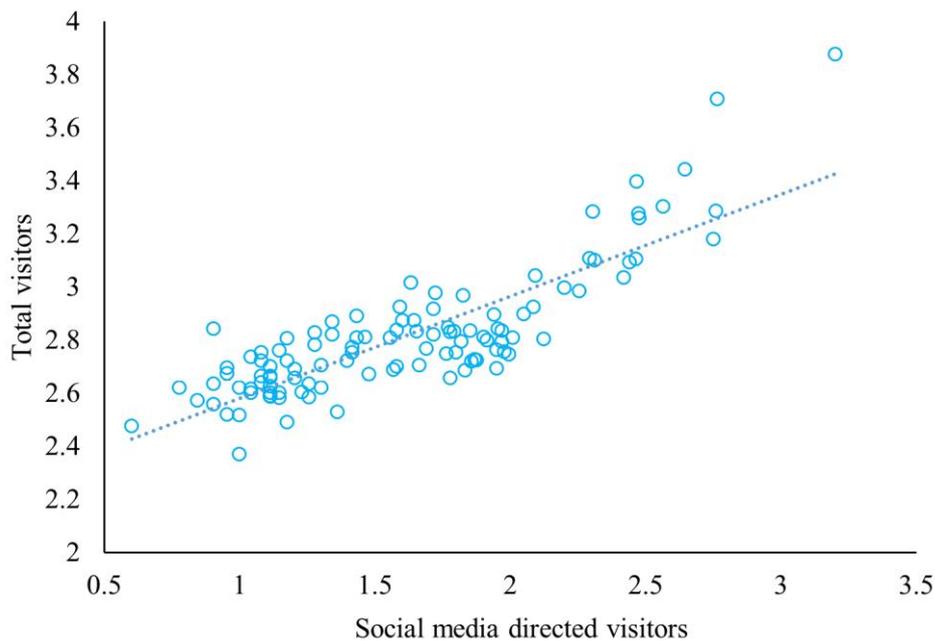

Figure 2 Scatter plot of social media (Log transformation)

*Correlation between Facebook shares/Tweets and visitors directed from Facebook/Twitter*

Facebook and Twitter are the two dominant social referrals directing people to scholarly articles, accounting for more than 95% of all social referrals. Individually Facebook and Twitter are roughly equivalent to one another (Wang, et al. 2016). Here the data of Facebook and Twitter are selected out and separated. In Figure 3, the blue dots represent the Twitter data, while the orange circles represent the Facebook data. The Y-axis corresponds to Facebook shares or Tweets, while the X-axis corresponds to the visitors directed from Facebook or Twitter. As Figure 3 shows, there is obvious stratification between the Twitter dots and Facebook circles. Compared with the Facebook circles, the Twitter dots are more closed to the horizontal axis, which indicates that compared with Facebook shares, Tweets directed more people to visit scholarly articles. Moreover, for Facebook, the correlation coefficient r = 0.854 (p<0.001); while for Twitter, the coefficient is 0.869 (p<0.001). Both correlations are significant.

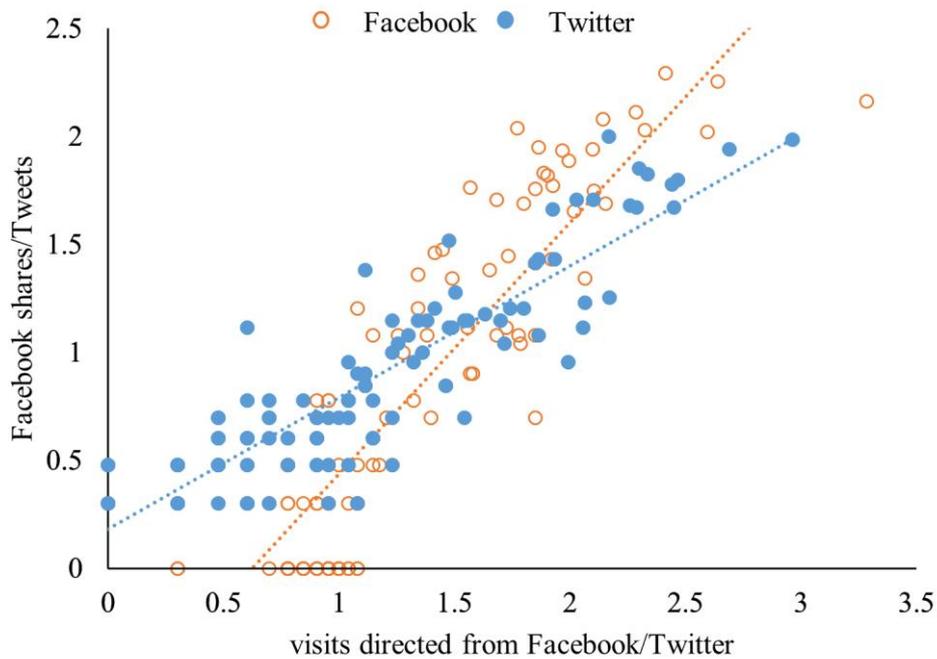

Figure 3 Scatter plot of social media attention and caused visits (Log transformation)

*Correlation between Facebook shares and Tweets*

For different social media platforms, do articles get equivalent attention? In other words, do articles receive more tweets also get more Facebook shares? To investigate this issue, we make correlation analysis of social media attention between Facebook and Twitter. Although the data in Figure 4 does not show a trend as obvious as Figure 2 and 3, it still indicates a positive relationship between Facebook shares and tweets.

According to the results of correlation analysis, Facebook shares are positively and strongly correlated with tweets, and the correlation coefficient r = 0.594 ($p<0.001$).

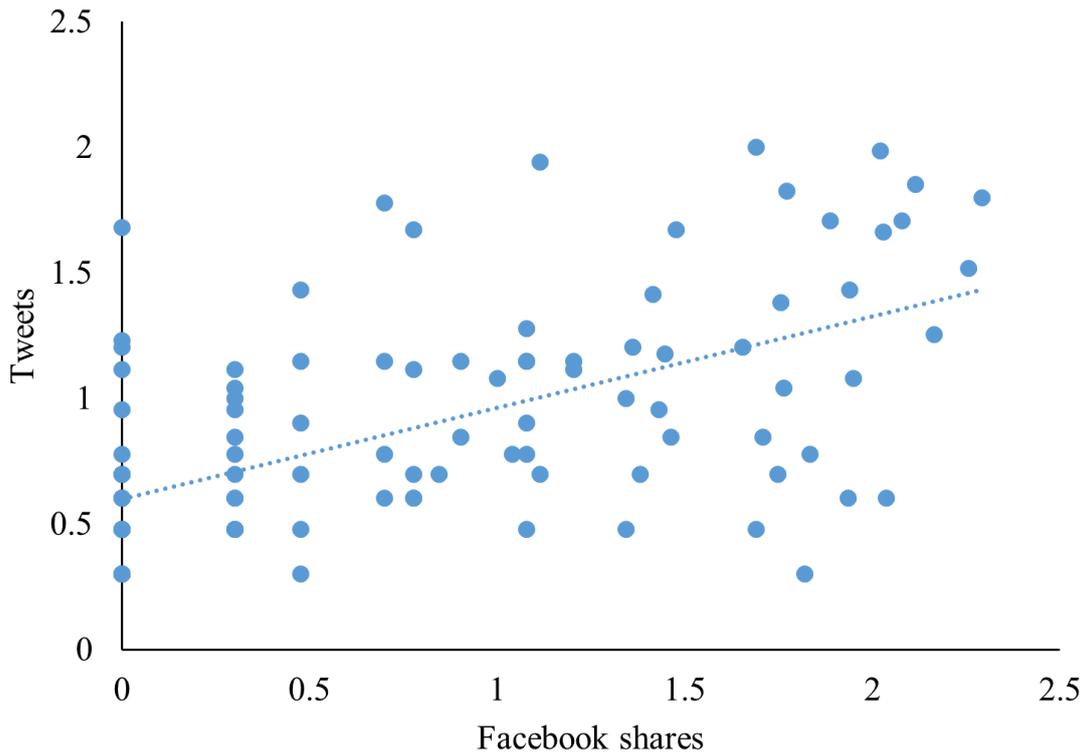

Figure 4 Correlation between Facebook shares and Tweets (Log transformation)

**3.3 Temporal trends**

For each paper, we record the tweeting time and calculate the interval days between tweeting and publishing. The tweets over time after publication show that most articles received tweets in a short time after their publication. Here we set a time point of 7 days, as Eysenbach (2011) did, and we calculate the total tweets (including tweets and retweets) within and after 7 days of article publication, correspondingly we count the Twitter directed visitors for each article in 7 days and after 7 days of publication. In Figure 5, we summarize the data for all articles in these two periods. 110 papers are tweeted 384 times in total, while papers got 95.27% of tweets in 7 days after publication, and only 5.73% of tweets are received in the later period. Twitter directed 5463 visitors to the 110 articles, while 72.30% of them came from the first 7 days after article publication.

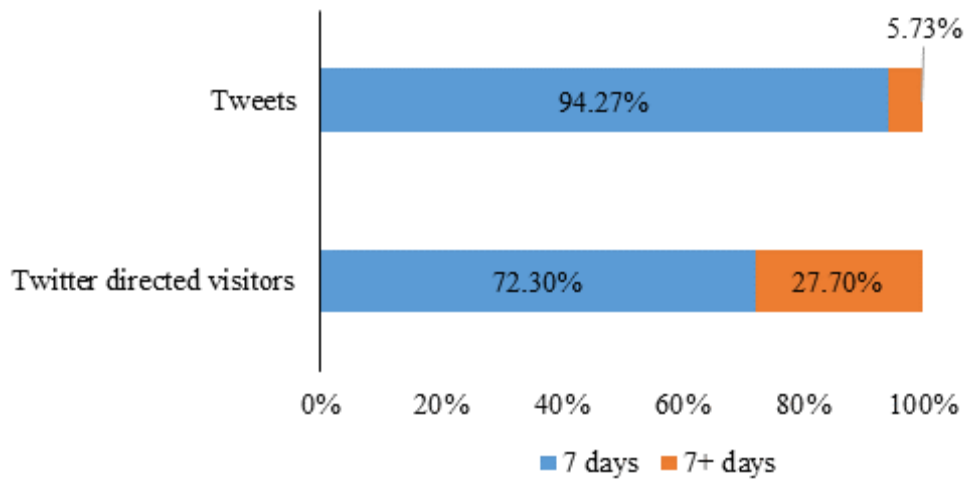

Figure 5 Distribution of total tweets and total Twitter directed visitors in 7 two periods after publication

One-way ANOVA is used to test whether there are significant differences in the number of tweets between the two periods, which are within 7 days and after 7 days of publication. Furthermore, we make the same analysis on the number of Twitter directed article visitors. The alpha level is set to 0.05. As shown in Table 3, the result is significant. The sig values of both tests are less than 0.05, which means that regardless of the number of tweets or the number of Twitter directed article visitors, there are significant differences between the number within 7 days and 7 days later.

Table 3 One-way ANOVA

|  | Mean | | F | Sig |
| --- | --- | --- | --- | --- |
|  | 7 days | 7 days later |  |  |
| Tweets | 11.7 | 2.6 | 30.064 | 0.000 |
| Visitors | 35.9 | 13.8 | 6.638 | 0.011 |

Figure 6 shows the statistics for each paper. Figure 6(a) indicates the tweets count, while Figure 6(c) is the enlargement of the top 10 papers with the most tweets; Figure 6(b) displays the visitor count, while Figure 6(d) is the enlargement of 10 papers corresponding to Figure 6(c). Each stacked bar represents the number of tweets/visitors for each article. The bar length is decided by the total number of tweets/visitors of the paper. The data in both panels are ranked by the total number of tweets for each paper. As Figure 6(a) shows, for most articles, the blue bar is much longer than the orange bar, which indicates that most articles received most tweets in the first 7 days after publication. Only one paper (https://doi.org/10.7717/peerj.1573) received more tweets in the late period (7+ days) than the early period (the first 7 days). Especially for the papers got a few tweets, almost all the tweets are received in the first 7 days. Figure 6(b) shows the Twitter directed visitors for each paper. Generally, articles with more tweets tend to have more visitors. However, there are also some exceptions. For example, paper 1573 (https://doi.org/10.7717/peerj.1573) has the most tweets, but with relative few visitors directed from Twitter.

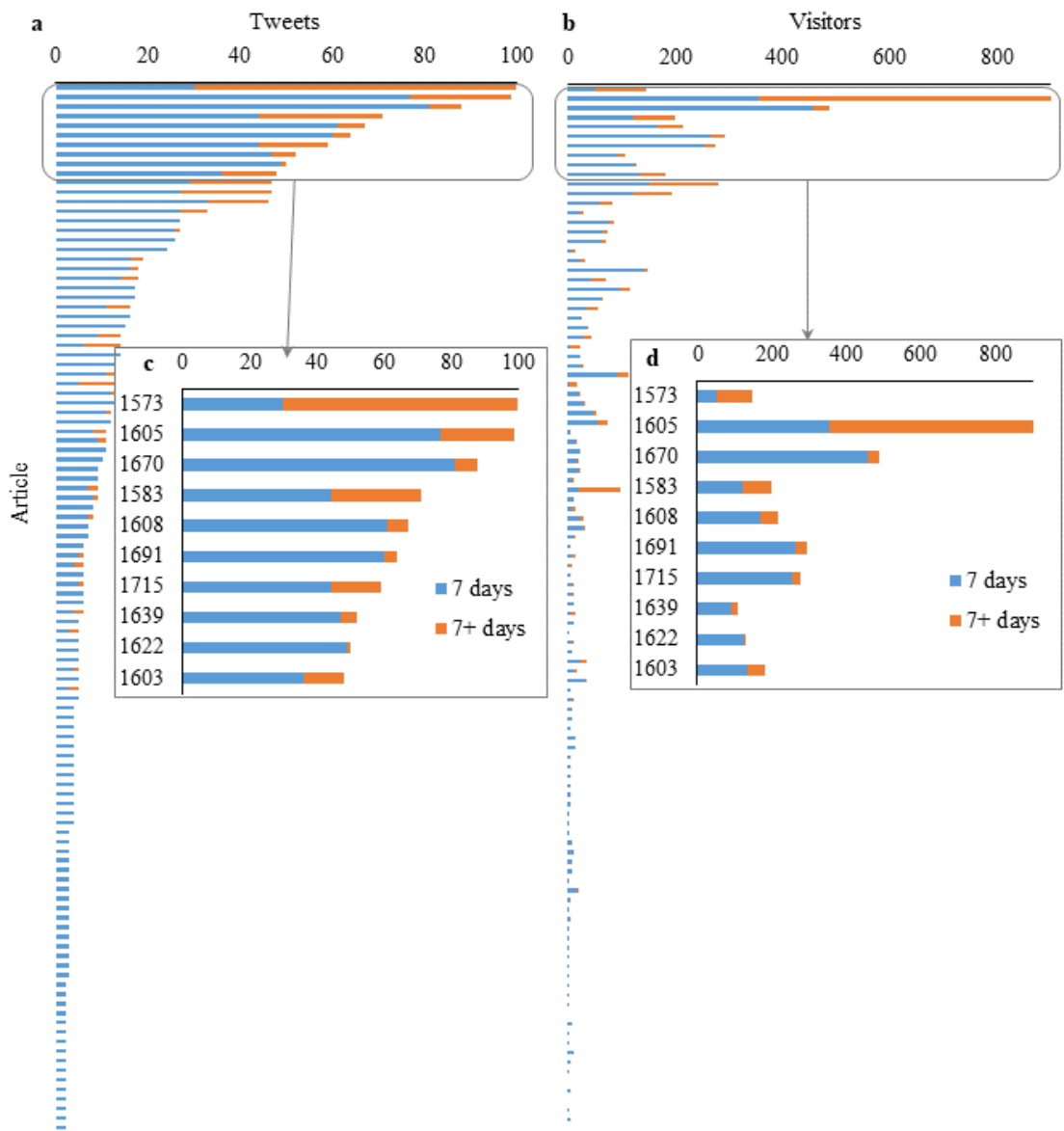

Figure 6 Distribution of tweets and Twitter directed visitors for each paper in two periods after publication

## 4. Conclusions and discussion

Firstly, social media attention increases the number of views of scholarly articles, which is confirmed by the direct evidence of social media directed article visitors. More social media attention suggests more article views, while some social media directed article visitors may not be reached through traditional ways. Secondly, there exist significant correlation among different social media activity. Articles with more Facebook shares tend to get more tweets, and vice versa. Thirdly, the temporal trends show that social attention comes immediately following publication. However, those coming easily may often go soon, social media attention around scholarly articles does not last long, the same applies to social media directed article views.

To better understand the role of social media in directing people to visit scholarly articles, this paper investigates the relationship between social media attention and article visitors at article-level. We employ unique referral data of 110 *PeerJ* articles, which could better illustrate the relationship between social media attention and social media directed visitors for each article. We record and analyze the daily updated visiting data of each article for a period of 345 days.

Our results show that the social media presence of PeerJ articles is high. About 68.18% of the papers received at least one tweet from Twitter accounts other than the official account of the journal.

Social media brings scholarly articles to the public. Not only researchers, but also many general people are directed to scholarly articles by social media attention. Although it needs more evidence to make deep and detailed analysis. Besides the complementary role to traditional, citation-based metrics (Priem, Piwowar, & Hemminger, 2012), online attention could be transformed to other kinds of impacts, e.g., article downloads. Social media attention increases the dissemination of scholarly articles. Scholarly articles attract visitors through their social media presence. Articles with more social media attention would have more article visitors. Social media directed visitors contribute significantly to the total article visitors, which is applicable for both Facebook and Twitter.

There also exist significant correlations among the online attention from different social media platforms. Articles with more Facebook shares tend to attract more tweets. It could be explained by the following reasons. Firstly, the article attracts independent users from Facebook or Twitter with no interference from the other to share it on social media platforms. Secondly, there may be overlapped user group across Facebook and Twitter. According to the report of Pew Research Center in 2013, 90% of Twitter users also use Facebook, and 22% of Facebook users also use Twitter (Duggan, & Smith, 2013). Article visitors directed by Twitter referral may share the paper on Facebook and vice versa.

The temporal trends show that social attention comes soon. Most of those tweets (94.27%) and Twitter directed visitors (72.30%) are concentrated in the few days immediately following publication, which are in consistent with the results of Eysenbach (2011), which find that the majority of tweets were sent within the 7 days of article publication, especially the day and the following day of article publication. Although we set the time window of 7 days in this study, we do observe tweets come earlier. The exact day and the following day of publication have the most tweets. However, those coming easily may often go soon, social media attention around scholarly articles does not last long. Only a few (5.73%) tweets distribute in the period from the 7$^{th}$ day to 345$^{th}$ day after publication, which generated 27.70% of all Twitter directed visitors.

There are some limitations in this study. Firstly, only 110 articles are included in the dataset, there exist sample size bias for the dataset. Secondly, besides the correlation between social media attention and social media directed visitors, the causality between the two factors maybe tell us more. Thirdly, we only collect the referrals data from PeerJ, which is a journal publishes articles in the specific field of life, biology and health science. There may also exists disciplinary bias. The universality of the findings needs to be examined in other disciplines. Moreover, there exist some disadvantages of altmetrics, including commercialization, data quality, missing evidence and manipulation (Bornmann, 2014), these shortcomings of altmetrics may have influence on the result.

## Acknowledgments

This research is supported by the "National Natural Science Foundation of China" (71673038, 61301227).